%% file: MorPiccMishRadi_arxiv2.2.tex
\documentclass[11pt]{article}

\usepackage{graphicx}
\usepackage{natbib}
\usepackage{setspace}
\usepackage{color}

\input macro

\begin{document}

\centerline{\Large \textbf{Integral identities for a semi-infinite interfacial crack}}

\centerline{\Large  \textbf{in anisotropic elastic bimaterials} }

\begin{center}
L. Morini$^{(1),(2)}$, A. Piccolroaz$^{(2),(3)}$, G. Mishuris$^{(3)}$ and E. Radi$^{(1)}$\\
\end{center}

\centerline{$^{(1)}$\emph{Dipartimento di Scienze e Metodi dell'Ingegneria, Universit\'a di Modena e Reggio Emilia,}}

\centerline{\emph{Via Amendola 2, 42100, Reggio Emilia, Italy.}}

\centerline{$^{(2)}$\emph{Department of Civil, Environmental and Mechanical Engineering, University of Trento,}}

\centerline{\emph{Via Mesiano 77, 38123, Trento, Italy.}}

\centerline{$^{(3)}$\emph{Institute of Mathematical and Physical Sciences, Aberystwyth University,}}

\centerline{\emph{ Ceredigion SY23 3BZ, Wales, U.K.}}

\begin{abstract}
The focus of the article is on the analysis of a semi-infinite crack at the interface between two dissimilar anisotropic elastic materials, loaded by a general asymmetrical system of forces acting on the crack faces. 
Recently derived symmetric and skew-symmetric weight function matrices are introduced for both plane strain and antiplane shear cracks, and used together with the fundamental reciprocal identity (Betti formula) in order 
to formulate the elastic fracture problem in terms of singular integral equations relating the applied loading and the resulting crack opening. The proposed compact formulation can be used to solve many problems in 
linear elastic fracture mechanics (for example various classic crack problems in homogeneous and heterogeneous anisotropic media, as piezoceramics or composite materials). This formulation is also fundamental in many 
multifield theories, where the elastic problem is coupled with other concurrent physical phenomena. 
\end{abstract}


\section{Introduction and formulation of the problem}

The method of singular integral equations in linear elasticity was first developed for solving two-dimensional problems, \citep{Muskh2}, and later extended to three-dimensional cases by means of 
multi-dimensional singular integral operators theory \citep{KuprGeg1, MikhPross1}. Singular integral formulations for both two and three-dimensional crack problems have been derived by means of a general approach based 
on Green's function method \citep{Weaver1, BudRic1, LinkZub1}. As a result, the displacements and the stresses are defined by integral relations involving the Green's functions, for which explicit expressions are 
required \citep{BigCap1}. 
Although Green's functions for many two and three-dimensional crack problems in isotropic and anisotropic elastic materials have been derived \citep {SinHirt1, Weaver1, Pan1, Pan2, PanYuan1}, their utilization in 
evaluating physical displacements and stress fields on the crack faces implies, especially in the anisotropic case, challenging numerical estimation of 
integrals which convergence should be asserted carefully. Moreover, the approach based on Green's function method works when the tractions applied on the discontinuity surface are symmetric, but not in the case of 
asymmetric loading acting on the crack faces.

Recently, using a procedure based on Betti's reciprocal theorem and weight functions\footnote{Defined by \citet{Bueck1} as singular non-trivial solutions of the homogeneous traction-free 
problem and later derived for general three-dimensional problems by \citet{Wilmov1}, and for interfacial cracks  by \citet{Gao1} and \citet{PiccMish1}.} an alternative method for deriving integral identities relating 
the applied loading and the resulting crack opening has been developed for two and three-dimensional semi-infinite interfacial cracks between dissimilar isotropic materials by \citet{PiccMish3}. In the 
two-dimensional case, the obtained identities contain Cauchy type singular operators together with algebraic terms. The algebraic terms vanish in the case of homogeneous materials. This 
approach avoids the use of the Green's functions without assuming the load to be symmetric. 

The aim of this paper is to derive analogous integral identities for the case of semi-infinite interfacial cracks in anisotropic bimaterials subjected to two-dimensional deformations. 

General expressions for symmetric and skew-symmetric weight functions for interfacial cracks in two-dimensional anisotropic bimaterials have been recently derived by \cite{MorRad1} by means of Stroh representation of 
displacements and fields \citep{Stroh1} combined with a Riemann-Hilbert formulation of the traction problem at the interface \citep{Suo1}. These expressions for the weight functions are used together with the results 
obtained for isotropic media by \citet{PiccMish3} in order to obtain integral formulation for interfacial cracks problems in anisotropic bimaterial solids with general asymmetric load applied at the crack faces.

\begin{figure}[!htcb]
\centering
\includegraphics[width=90mm]{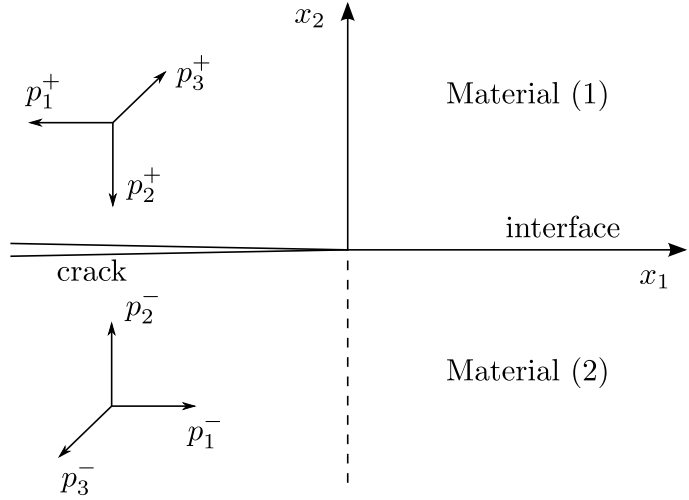}
\caption[geom_frac_id]{\footnotesize {Two-dimensional semi-infinite interfacial crack loaded by non necessarily symmetric forces applied on the crack faces.}}
\label{geom_frac_id}
\end{figure}

We consider a two-dimensional semi-infinite crack between two dissimilar anisotropic elastic materials with asymmetric loading applied to the crack faces, the geometry of the system is shown in Fig.\ref{geom_frac_id}. 
Further in the text, we will use the superscripts $^{(1)}$ and $^{(2)}$ to denote the quantities related to the upper and the lower elastic half planes, respectively. The crack is situated along the negative semi-axis 
$x_{1} < 0$. Both in-plane and antiplane stress and deformation, which in fully anisotropic materials are coupled \citep{Ting2}, are taken into account. The symmetrical and skew-symmetrical 
parts of the loading are defined as follows:
\beq
\left\langle \Bp\right\rangle = \fr{1}{2}\left(\Bp^{+} + \Bp^{-}\right),\quad \jump{0.1}{\Bp} = \Bp^{+} - \Bp^{-},
\eequ{p_symskew}
where $\Bp^+$ and $\Bp^-$ denote the loading applied on the upper and lower crack faces, $x_2 = 0^+$ and $x_2 = 0^-$, respectively (see Fig. \ref{geom_frac_id}).

In Section \ref{Preliminary} preliminary results needed for the derivation of the integral identities and for the complete explanation of the proposed method are reported. In Section \ref{Betti_weight}, the fundamental 
reciprocal identity and the weight functions, defined as special singular solution of the homogeneous traction-free problem are introduced. In Section \ref{Weight}, symmetric and skew-symmetric weight functions matrices 
for interfacial cracks in anisotropic bimaterials recently derived by \citet{MorRad1} are reported.

Section \ref{Monoclinic} contains the main results of the paper: integral identities \eq{int_anti1S}, \eq{int_anti2S}, \eq{int_plane1S} and \eq{int_plane2S} for two-dimensional crack problems between two dissimilar 
anisotropic materials are derived and discussed in details. The integral identities are derived for monoclinic-type materials, which are the most general class of anisotropic media where both in-plane 
and antiplane strain and in-plane and antiplane stress are uncoupled \citep{Ting2,Ting3}, and the Mode III can be treated separately by Mode I and II. By means of Betti's formula and weight functions, 
both antiplane and plane strain fracture problems are formulated in terms of singular integral equations relating the applied loading and the resulting crack opening. 

In Section \ref{examples}, the obtained integral identities are used for studying cracks in monoclinic bimaterials loaded by systems of line forces acting on the crack faces. The proposed examples show that using the 
identities explicit expressions for crack opening and tractions ahead of the crack tip can be derived for both antiplane and in-plane problems. These simple illustrative cases demonstrate also that the 
proposed integral formulation is particularly easy to apply and can be very useful especially in the analysis of phenomena where the elastic behaviour of the material is coupled with other physical effects, as for example 
hydraulic fracturing, where both anisotropy of the geological materials and fluid motion must be taken into account.

Finally, in Appendix A, the Stroh formalism \citep{Stroh1}, adopted by \cite{Suo1} and \cite{GaoAbbu1} in analysis of interfacial cracks in anisotropic bimaterials and recently used by \cite{MorRad1} for deriving 
symmetric and skew-symmetric weight functions, is briefly explained. In particular, explicit expressions for Stroh matrices and surface admittance tensor needed in weight functions expressions associated to monoclinic 
materials are reported.

\section{Preliminary results}
\label{Preliminary}

In this section relevant results obtained by several studies regarding interfacial cracks are reported. These results will be used further in the paper in order to develop an integral formulation for the problem of a 
semi-infinite interfacial crack in anisotropic bimaterials. 

In Section \ref{Betti_weight}, we introduce the Betti integral formula for a crack in an elastic body subjected to two-dimensional deformations with general asymmetric loading applied at the faces.

In Section \ref{Weight}, general matrix equations expressing weight functions in terms of the associate singular traction vectors, recently derived by \citet{MorRad1}, and valid for interfacial cracks 
in a wide range of two-dimensional anisotropic bimaterials are reported.

\subsection{The Betti formula}
\label{Betti_weight}

The Betti formula is generally used in linear elasticity in order to relate the physical solution to the weight function which is defined as special singular solution to the homogeneous traction-free problem 
\citep{Bueck1, Wilmov1}. Since the Betti integral theorem is independent of the specific elastic constitutive relations of the material, it applies to both isotropic and anisotropic 
media in the same 
form.

The notations $\Bu=(u_1,u_2,u_3)^T$ and $\BGs=(\Gs_{21},\Gs_{22},\Gs_{23})^T$ are introduced to indicate respectively the physical displacements and the traction vector acting on the plane $x_2=0$. According to the fact 
that two-dimensional elastic deformations are here considered, both displacements and stress do not depend on the variable $x_3$. Nevertheless, since both in-plane and anti-plane strain and stress are considered, non-
zero components $u_3$ and $\sigma_{23}$ are accounted for \citep{Ting2}. The notations $\BU=(U_1,U_2,U_3)^T$ and $\BGS=(\GS_{21},\GS_{22},\GS_{23})^T$ are introduced to indicate the weight function, defined by 
\citet{Bueck1} as a non-trivial singular solution of the homogeneous traction-free problem, and the associated traction vector, respectively.  As it was shown by \citet{Wilmov1}, the weight function $\BU$ is defined in a 
different domain respect to physical displacement, where the crack is placed along the positive semi-axis $x_2>0$. Following the procedure reported and 
discussed in \citet{Wilmov1,PiccMish1} and \citet{PiccMish3}, from the application of the Betti integral formula to the physical fields and to weight functions for both the upper and the lower half-planes in Fig.
\ref{geom_frac_id}, we obtain:
$$
\int_{-\infty}^{\infty} \left\{\tilde{\BR}\BU(x_{1}'-x_1,0^+) \cdot \Bp^{+}(x_1) - \tilde{\BR}\BU(x_{1}'-x_1,0^-) \cdot \Bp^{-}(x_1)+\right.
$$
$$
\left.+\tilde{\BR}\BU(x_{1}'-x_1,0^+)\cdot\BGs^{(+)}(x_1,0^+)-\tilde{\BR}\BU(x_{1}'-x_1,0^-)\cdot\BGs^{(+)}(x_1,0^-)-\right.
$$
\beq
\left.-\left[\tilde{\BR}\BGS(x_{1}'-x_1,0^+)\cdot\Bu(x_1,0^+)+\tilde{\BR}\BGS(x_{1}'-x_1,0^-)\cdot\Bu(x_1,0^-)\right]\right\}dx_1=0 ,
\eequ{betti1}
where $\tilde{\BR}$ is the rotation matrix:
$$
\tilde{\BR} = \pmatrix{-1 & 0  & 0 \cr 0 & 1 & 0 \cr 0 & 0 & -1},
$$
$\Bp^{+}(x_1)$ and $\Bp^{-}(x_1)$ are the loading acting on the upper and on the lower crack faces, respectively, and $\BGs^{(+)}(x_1,0)$ is the physical traction at the interface 
ahead of the crack tip. The superscript $^{(+)}$ denotes a function whose support is restricted to the positive semi-axis, $x_1 > 0$. In eq. (\ref{betti1}), $x_1'$ denotes a shift of 
the weight function within the plane $(x_1,x_3)$ and the dot symbol stands for the scalar product.

Assuming perfect contact conditions at the interface, which implies displacement and traction continuity at the interface ahead of the crack tip, $\BGs^{(+)}(x_1,0)$ can be defined as follows:
\beq
\BGs^{(+)}(x_1,0^+)=\BGs^{(+)}(x_1,0^-)=\bfm\tau^{(+)}(x_1), \quad x_1>0.
\eequ{taudef}
Similarly, also the traction corresponding to the singular solution $\BU$ satisfies the continuity at the interface:
\beq
\BGS(x_1,0^+)=\BGS(x_1,0^-)=\BGS(x_1), \quad x_1<0. 
\eequ{Sigmadef}
Using these definitions, \eq{betti1} becomes:
\begin{displaymath}
\int_{-\infty}^{\infty} \left\{\tilde{\BR}\jump{0.1}{\BU}(x_1'-x_{1}) \cdot \bfm\tau^{(+)}(x_{1}) - \tilde{\BR}\BGS(x_{1}'-x_1) \cdot \jump{0.1}{\Bu}^{(-)}(x_1)\right\}dx_1 = 
\end{displaymath}
\beq
= -\int_{-\infty}^{\infty} \left\{\tilde{\BR}\BU(x_{1}'-x_1,0^+) \cdot \Bp^{+}(x_1) - 
\tilde{\BR}\BU(x_{1}'-x_1,0^-) \cdot \Bp^{-}(x_1)\right\}dx_1,
\eequ{betti}
where $\jump{0.1}{\Bu}^{(-)}$ is the crack opening behind the tip, $^{(-)}$ denotes that its support is restricted to the negative semi-axis, $x_1 < 0$, and $\jump{0.1}{\BU}$ is known as the symmetric 
weight function \citep{Wilmov1,PiccMish1, MorRad1}:
\beq
\jump{0.1}{\BU}(x_1)=\BU(x_1,0^+)-\BU(x_1,0^-).
\eequ{wf_sym}
By expressing the loading acting on the crack faces in terms of the symmetric and skew-symmetric parts defined by \eq{p_symskew}, the Betti identity \eq{betti} finally becomes:
\begin{displaymath}
\int_{-\infty}^{\infty} \left\{\tilde{\BR}\jump{0.1}{\BU}(x_1'-x_{1})\cdot\bfm\tau^{(+)}(x_{1})-\tilde{\BR}\BGS(x_{1}'-x_1)\cdot\jump{0.1}{\Bu}^{(-)}(x_1)\right\}dx_1 = 
\end{displaymath}
\beq
= -\int_{-\infty}^{\infty} \left\{\tilde{\BR}\jump{0.1}{\BU}(x_{1}'-x_1)\cdot\left\langle \Bp\right\rangle(x_1) + 
\tilde{\BR}\left\langle \BU\right\rangle(x_{1}'-x_1)\cdot\jump{0.1}{\Bp}(x_1)\right\}dx_1,
\eequ{betti_final}
where $\left\langle \BU\right\rangle$ is known as the skew-symmetric weight function \citep{Wilmov1,PiccMish1,  MorRad1}:
\beq
\left\langle \BU\right\rangle(x_1)=\fr{1}{2}\left[\BU(x_1,0^+)+\BU(x_1,0^-)\right].
\eequ{wf_skew}
The integral identity \eq{betti_final} can be written in an equivalent form using the convolution respect to $x_1$, denoted by the symbol $*$ \citep{Arfk1}:
\beq
(\tilde{\BR}\jump{0.1}{\BU})^{T} * \bfm\tau^{(+)} - (\tilde{\BR}\BGS)^{T} * \jump{0.1}{\Bu}^{(-)} = 
-(\tilde{\BR}\jump{0.1}{\BU})^{T} * \langle\Bp\rangle - (\tilde{\BR}\langle\BU\rangle)^{T} * \jump{0.1}{\Bp}.
\eequ{Betti_conv}
This integral identity relates physical traction and crack opening to weight functions and load applied at the crack faces, and will be used further in the text in order to formulate the interfacial crack problem 
between dissimilar anisotropic materials in terms of singular integral equations. 

Note that, in order to simplify notations, in eq. (\ref{Betti_conv}) the scalar product $\Ba \cdot \Bb$ between vectors $\Ba$ and $\Bb$ is replaced by the ``row by column'' product $\Ba^T \Bb$ between 
the row vector $\Ba^T$ and the column vector $\Bb$.

\subsection{Symmetric and skew-symmetric weight functions for anisotropic bimaterials}
\label{Weight}

Let us introduce the Fourier transform of a generic function $f$ with respect to the variable $x_1$ as follows:
\beq
\hat{f}(\xi) = \CF_{\xi}[f(x_1)] = \int_{-\infty}^\infty f(x_1) e^{i \xi x_{1}} dx_{1}, \quad 
f(x_1) = \CF^{-1}_{x_1}[\hat{f}(\xi)] = \frac{1}{2\pi} \int_{-\infty}^\infty \hat{f}(\xi) e^{-i \xi x_{1}} d\xi.
\eequ{fourier}
 In \citet{MorRad1}, the following expressions for the Fourier transform of the singular displacements $\BU$ at the interface between two dissimilar anisotropic media have been derived:
\beq
\hat{\BU} (\xi,0^+) = \Big\{  \fr{1}{2 \xi} (\BY^{(1)} - \ov{\BY}^{(1)}) - \fr{1}{2 |\xi |} (\BY^{(1)} + \ov{\BY}^{(1)}) \Big\} \hat{\BGS}^-(\xi),\quad \xi\in\textrm{R},
\eequ{U+}
\beq
\hat{\BU} (\xi,0^-) = \Big\{  \fr{1}{2 \xi} (\BY^{(2)} - \ov{\BY}^{(2)}) + \fr{1}{2 |\xi |} (\BY^{(2)} + \ov{\BY}^{(2)}) \Big\} \hat{\BGS}^-(\xi),\quad \xi\in\textrm{R}, 
\eequ{U-}
where $\hat{\BGS}^-$ is the Fourier transform of the singular traction at the interface, which in the case of perfect contact condition is defined as in expression \eq{Sigmadef},
and $\BY$ is the Hermitian definite positive surface admittance tensor \citep{GaoAbbu1}, depending on the elastic properties of the materials and defined in details in Appendix A.

The superscripts $^+$ and $^-$, used here and in the sequel, denote functions analytic in the upper and in the lower complex half-planes, respectively
$$
\hat{f}^+(\xi) = \CF_{\xi}[f^{(+)}(x_1)], \quad \hat{f}^-(\xi) = \CF_{\xi}[f^{(-)}(x_1)].
$$

Eqs. (\ref{U+}) and (\ref{U-}) represent general expressions relating the tractions $\hat{\BGS}^-(\xi)$ applied on the bounding surfaces and the corresponding displacements $\hat{\BU} (\xi,0^+)$, 
$\hat{\BU} (\xi,0^-)$ for the upper and lower half-planes, respectively.

The symmetric and skew-symmetric weight function matrices are derived by taking respectively the jump and the average of $\hat{\BU}$ \citep{Wilmov1,PiccMish1}:
\beq
\jump{0.1}{\hat{\BU}}^+(\xi)= 
\fr{1}{|\xi|} \Big\{ i ~ \mbox{sign} (\xi) ~ \mbox{Im} (\BY^{(1)}- \BY^{(2)})  - \mbox{Re} (\BY^{(1)} + \BY^{(2)})   \Big\} \hat{\BGS}^-(\xi) ,
\eequ{Usym}
\beq
\langle \hat{\BU} \rangle(\xi) = \fr{1}{2 |  \xi |} \Big\{ i ~ \mbox{sign} (\xi) ~ \mbox{Im} (\BY^{(1)}+ \BY^{(2)})  - \mbox{Re} (\BY^{(1)} - \BY^{(2)})   \Big\} \hat{\BGS}^-(\xi),\quad \xi\in\textrm{R} .
\eequ{Uskew}
Eqs. \eq{Usym} and \eq{Uskew} can also be expressed in the compact form:
\beq
\jump{0.1}{\hat{\BU}}^+(\xi)= 
-\fr{1}{|\xi|} \Big\{ \mbox{Re}\BH-i ~ \mbox{sign} (\xi) ~ \mbox{Im}\BH   \Big\} \hat{\BGS}^-(\xi) ,
\eequ{UsymH}
\beq
\langle \hat{\BU} \rangle(\xi) = -\fr{1}{2 |  \xi |} \Big\{\mbox{Re} \BW- i ~ \mbox{sign} (\xi) ~ \mbox{Im}\BW  \Big\} \hat{\BGS}^-(\xi),\quad \xi\in\textrm{R},
\eequ{UskewW}
where $\BH$ and $\BW$ are the bimaterial matrices defined as follows \citep{Suo1,Ting3}:
\begin{eqnarray}
\BH & = & \BY^{(1)} + \ov{\BY}^{(2)}\label{H},\\
\BW & = & \BY^{(1)} - \ov{\BY}^{(2)}\label{W}.
\end{eqnarray}
Expressions \eq{UsymH} and \eq{UskewW} are valid for interfacial cracks in general anisotropic two-dimensional media. Since in anisotropic materials in-plane and antiplane displacements and stresses are 
generally coupled \citep{Ting1,Ting3}, for the case of fully anisotropic media three linearly independent vectors $\BGS$ and then $\BU$ must be defined for obtaining a complete basis of the singular solutions space. 
Nevertheless, there are several classes of anisotropic materials where in-plane and antiplane displacements and stress are uncoupled \citep{Ting2,Ting3} and then Mode III deformation can be treated 
separately from Mode I and II as for the case of isotropic media \citep{PiccMish1,PiccMish4,PiccMish3}. In the next Section integral identities are derived for interfacial crack between two media belonging to the most 
general of these classes, known as monoclinic materials.

\section{Integral identities}
\label{Monoclinic}

In this Section, following the approach of \citet{PiccMish3}, an integral formulation of the problem of a semi-infinite two-dimensional interfacial crack in anisotropic bimaterials is obtained. A 
particular class of anisotropic materials, where elastic properties are symmetrical with respect to a plane, is considered. These materials are known as monoclinic, and in the case in which the plane of symmetry 
coincides with $x_3=0$ both in-plane and antiplane displacements and in-plane and antiplane stress are uncoupled \citep{Ting2}. Monoclinic having plane of symmetry at 
$x_3=0$ are the most general class of anisotropic materials where stress and strain are decoupled, and they include as subgroups all other classes having this property, such as orthotropic and cubic materials 
\citep{HorMil1,Ting3}. Explicit expressions for Stroh matrices and surface admittance tensor corresponding to these type of media are reported in Appendix A. These expressions have been used for evaluating bimaterial 
matrices (\ref{H}) and (\ref{W}).

In Sections \ref{antiplane_mon} and \ref{inplane_mon}, antiplane shear and plane strain interfacial cracks problems in monoclinic bimaterials are formulated in terms of singular integral equations by means of weight 
function expressions \eq{UsymH} and \eq{UskewW} and Betti integral identity \eq{Betti_conv}.

\subsection{Mode III}
\label{antiplane_mon}

Considering antiplane deformations in monoclinic materials, as it is shown in Appendix A, constitutive relations reduce to scalar equations relating stresses $\Gs_{23}$ and $\Gs_{13}$ to $u_3$, and then the traction 
$\Gs_{23}=\tau^{+}_{3}$ and the displacements derivative for both upper and lower half-plane material become \citep{Suo1}:
\beq
\tau_3(x_{1},x_{2}) =L_{33} h_3(z_3) + \ov{L_{33} h_3(z_3)},
 \eequ{trac_anti}
 \beq
 u_{3,1} (x_{1},x_{2}) = F_{33} h_3(z_3) + \ov{F_{33} h_3(z_3)},
 \eequ{displ_anti}
 where $z_3=x_1+\mu_3x_2$. The bimaterial matrices (\ref{H}) and (\ref{W}) reduce to:
\begin{eqnarray}
H_{33} & = &\left[\sqrt{s_{44}^{'}s_{55}^{'}-s_{45}^{'2}}\right]^{(1)}+\left[\sqrt{s_{44}^{'}s_{55}^{'}-s_{45}^{'2}}\right]^{(2)},\\
W_{33} & = & \left[\sqrt{s_{44}^{'}s_{55}^{'}-s_{45}^{'2}}\right]^{(1)}-\left[\sqrt{s_{44}^{'}s_{55}^{'}-s_{45}^{'2}}\right]^{(2)},
\end{eqnarray}
where $s_{ij}^{'}$ are elements of the reduced elastic compliance matrix (see Appendix A). According to general expressions \eq{UsymH} and \eq{UskewW}, the Fourier transform of symmetric and skew-symmetric weight 
functions for an antiplane shear crack between two dissimilar monoclinic materials are:
\beq
\jump{0.1}{\hat{U}_{3}}^+(\xi)=-\fr{H_{33}}{|\xi|}\hat{\GS}_{23}^-(\xi),\quad \langle \hat{U}_{3} \rangle(\xi) = -\fr{W_{33}}{2|\xi|}\hat{\GS}_{23}^-(\xi)=\fr{\nu}{2}\jump{0.1}{\hat{U}_{3}}^+(\xi),
\eequ{weight_anti}
where the following non-dimensional parameter has been introduced:
\beq
\nu = \fr{\left[\sqrt{s_{44}^{'}s_{55}^{'}-s_{45}^{'2}}\right]^{(1)}-\left[\sqrt{s_{44}^{'}s_{55}^{'}-s_{45}^{'2}}\right]^{(2)}}
{\left[\sqrt{s_{44}^{'}s_{55}^{'}-s_{45}^{'2}}\right]^{(1)}+\left[\sqrt{s_{44}^{'}s_{55}^{'}-s_{45}^{'2}}\right]^{(2)}}.
\eequ{nu}
In the case of antiplane deformations, the Betti formula reduces to the scalar equation: 
\beq
\jump{0.1}{U_3}*\tau_{3}^{(+)}-\GS_{23}*\jump{0.1}{u_3}^{(-)}=-\jump{0.1}{U_3}*\left\langle p_3\right\rangle-\left\langle U_3\right\rangle*\jump{0.1}{p_3}.
\eequ{betti_anti}
Applying the Fourier transform with respect to $x_1$, defined by relation \eq{fourier}, to this identity, we obtain:
\beq
\jump{0.1}{\hat{U}_{3}}^+\hat{\tau}_{3}^{+}-\hat{\GS}_{23}^-\jump{0.1}{\hat{u}_3}^{-} = -\jump{0.1}{\hat{U}_{3}}^+\left\langle\hat{p}_3\right\rangle-\langle \hat{U}_{3} \rangle\jump{0.1}{\hat{p}_3}.
\eequ{betti_anti_fourier}
Multiplying both sides of \eq{betti_anti_fourier} by $\jump{0.1}{\hat{U}_{3}}^{-1}$, we obtain:
\beq
\hat{\tau}_{3}^{+}-B\jump{0.1}{\hat{u}_3}^{-}=-\left\langle\hat{p}_3\right\rangle-A\jump{0.1}{\hat{p}_3}.
\eequ{int_anti_fourier}
The factors in front of the unknown functions are given by:
\beq
A=\jump{0.1}{\hat{U}_{3}}^{-1}\langle \hat{U}_{3} \rangle=\fr{\nu}{2}, \quad B=\jump{0.1}{\hat{U}_{3}}^{-1}\hat{\GS}_{3}=-\fr{|\xi|}{H_{33}}.
\eequ{AB_anti}

If we apply the inverse Fourier transform to \eq{int_anti_fourier}, we derive two distinct relationships corresponding to the two cases $x_1<0$ and $x_1>0$:
\beq
\left\langle p_3 \right\rangle(x_1) + \CF^{-1}_{x_{1}<0}\Big[A\jump{0.1}{\hat{p}_3}\Big]=\CF^{-1}_{x_{1}<0}\Big[B\jump{0.1}{\hat{u}_3}^{-}\Big], \quad x_1 < 0,
\eequ{int_anti1}
\beq
\tau_{3}^{+}(x_1) = \CF^{-1}_{x_{1}>0}\Big[B\jump{0.1}{\hat{u}_3}^{-}\Big], \quad x_1 > 0.
\eequ{int_anti2}
It is important to note that the term $\hat{\tau}_{3}^{+}$ cancels from \eq{int_anti1} because it is a ``${+}$'' function, while $[\hat{p}_3]$ and $\left\langle\hat{p}_3\right\rangle$ cancel from \eq{int_anti2} because 
they are ``${-}$'' functions.

To proceed further, we need to evaluate the inverse Fourier transform of the function $|\xi|\jump{0.1}{\hat{u}_3}^{-}$. Following the procedure illustrated by \citet{PiccMish3}, we get:
\beq
\CF^{-1}_{x_1}\Big[|\xi|\jump{0.1}{\hat{u}_3}^{-}\Big] = 
\fr{1}{\pi x_1}*\fr{\partial \jump{0.1}{u_3}^{-}}{\partial x_1} = 
\fr{1}{\pi}\int_{-\infty}^\infty \fr{1}{x_1-\eta}\fr{\partial \jump{0.1}{u_3}^{-}}{\partial \eta} d\eta.
\eequ{conv_prod}
Then we can define the singular operator $\CS$ and the orthogonal projectors $\CP_{\pm}$ $(\CP_+ + \CP_-=\CI)$ acting on the real axis:
\beq
\psi=\CS\varphi=\fr{1}{\pi x_1}*\varphi(x_1)=\fr{1}{\pi}\int_{-\infty}^\infty \fr{\varphi(\eta)}{x_1-\eta} d\eta,
\eequ{S_op}
\beq
\CP_{\pm}\varphi=\left\{
\begin{array}{rl}
\varphi(x_1), & \quad\pm x_1\geq 0, \\
0           , & \quad\mbox{otherwise}.
\end{array}
\right.
\eequ{P_op}
The operator $\CS$ is a singular operator of Cauchy type, and it transforms any function $\varphi$ satisfying the H$\ddot{\mbox{o}}$lder condition into a new function $\CS\varphi$ which also satisfies this condition 
\citep{Muskh1}. The properties of the operator $\CS$ in several functional spaces have been described in details in \citet{Pross1}.

The integral identities \eq{int_anti1} and \eq{int_anti2} for a Mode III interfacial crack between two dissimilar monoclinic materials become:
\beq
\left\langle p_3\right\rangle(x_1) + \fr{\nu}{2}\jump{0.1}{p_3}(x_1) = -\fr{1}{H_{33}}\CS^{(s)}\fr{\partial \jump{0.1}{u_3}^{(-)}}{\partial x_1}, \quad x_1<0,
\eequ{int_anti1S}
\beq
\tau_{3}^{(+)}(x_1) = -\fr{1}{H_{33}}\CS^{(c)}\fr{\partial \jump{0.1}{u_3}^{(-)}}{\partial x_1}, \quad x_1>0,
\eequ{int_anti2S}
where $\CS^{(s)}=\CP_-\CS\CP_-$ is a singular integral operator, and $\CS^{(c)}=\CP_+\CS\CP_-$ is a compact integral operator \citep{Gakh1,Krein1,GohKr1}. These two operators look similar, but they are essentially 
different, in fact: $\CS^{(s)}: F(\textrm{R}_-)\rightarrow F(\textrm{R}_-)$, while $\CS^{(c)}: F(\textrm{R}_-)\rightarrow F(\textrm{R}_+)$, where $F(\textrm{R}_{\pm})$ is some functional space of functions defined on 
$\textrm{R}_{\pm}$.

For explaining better this point, the integral identities \eq{int_anti1S} and \eq{int_anti2S} can be written in the extended form:
\beq
\left\langle p_3\right\rangle(x_1) + \fr{\nu}{2}\jump{0.1}{p_3}(x_1) = -\fr{1}{\pi H_{33}}\int_{-\infty}^0 \fr{1}{x_1-\eta}\fr{\partial \jump{0.1}{u_3}^{(-)}}{\partial \eta} d\eta, \quad x_1<0,
\eequ{intid_anti1}
\beq
\tau_{3}^{(+)}(x_1) = -\fr{1}{\pi H_{33}}\int_{-\infty}^0 \fr{1}{x_1-\eta}\fr{\partial \jump{0.1}{u_3}^{(-)}}{\partial \eta} d\eta, \quad x_1>0.
\eequ{intid_anti2}
The integral in \eq{intid_anti1} is a Cauchy-type singular integral with a moving singularity, whereas the integral in \eq{intid_anti2} possesses a fixed point singularity \citep{Dud1,Dud2}.

In the case of a homogeneous monoclinic material, the integral identities \eq{int_anti1S} and \eq{int_anti2S} simplify, since $\nu=0$, $H_{33}=2\sqrt{s_{44}^{'}s_{55}^{'}-s_{45}^{'2}}$
and thus there is no influence of the skew-symmetric loading.

Summarizing, the integral identities for Mode III interfacial cracks in monoclinic bimaterials are given by equations \eq{int_anti1S} and \eq{int_anti2S}. The equation \eq{int_anti1S} in an invertible 
singular integral relation between the applied loading $\left\langle p_3\right\rangle$, $\jump{0.1}{p_3}$ and the corresponding crack opening $\jump{0.1}{u_3}^{(-)}$. The equation \eq{int_anti2S} is an additional 
relation through which it is possible to define the behaviour of the solution $\jump{0.1}{u_3}^{(-)}$. Since the operator $\CS^{(c)}$ is compact, it is not invertible, and thus for deriving the traction ahead of the 
crack tip $\tau_{3}^{(+)}$ one needs to evaluate $\jump{0.1}{u_3}^{(-)}$ by inversion of the equation \eq{int_anti1S} (see \citet{Muskh1} for details).

\subsection{Mode I and II}
\label{inplane_mon}

For plane strain deformations in monoclinic materials, the surface admittance tensor $\BY$ is given by a $2\times2$ matrix of the form \citep{Ting2}:
\beq
\BY=s_{11}^{'}\BP+i(s_{11}^{'}c-s_{12}^{'})\BE,
\eequ{Ymatrix_plane}
where:
\beq
\BP = \pmatrix{b & d \cr d & e}, \quad \BE = \pmatrix{0 & -1 \cr 1 & 0},
\eequ{PE_matrices}
\beq
\mu_1+\mu_2=a+ib, \quad \mu_1\mu_2=c+id,
\eequ{mu_sumprod}
\beq
e=ad-bc=\mbox{Im}[\mu_1\mu_2(\ov{\mu}_1+\ov{\mu}_2)],
\eequ{mu_inv}
in which $\mu_1$ and $\mu_2$ are solutions of the eigenvalue problem associated to balance equations by means of Stroh representation of displacements and stresses 
\citep{Stroh1, Ting1} (see Appendix A for more details). Since $\mu_1$ and $\mu_2$ are eigenvalues with positive imaginary part, $b$ is strictly positive, $b=\mbox{Im}(\mu_1+\mu_2)>0,$ while 
the positive definiteness of the matrix $\BY$, and consequently of $\BP$, implies that \citep{Ting2}:
\beq
e>0 \quad \mbox{and} \quad be-d^2<0.
\eequ{Dund_lim}
Thus, bimaterial matrices $\BH$ and $\BW$ for an interfacial crack between dissimilar monoclinic materials under plane strain deformations can be decomposed into real and imaginary 
parts as follows:
\beq
\BH=\BY^{(1)}+\ov{\BY}^{(2)}=\BH^{'}+i\beta\sqrt{H_{11}H_{22}}\BE,
\eequ{H_matrice}
\beq
\BW=\BY^{(1)}-\ov{\BY}^{(2)}=\BW^{'}-i\gamma\sqrt{H_{11}H_{22}}\BE,
\eequ{W_matrice}
where matrices $\BH^{'}$ and $\BW^{'}$ are defined as:
\beq
\BH^{'} = \pmatrix{H_{11} & \Ga\sqrt{H_{11}H_{22}} \cr \Ga\sqrt{H_{11}H_{22}} & H_{22}}, \quad \BW^{'} = \pmatrix{\delta_1 H_{11}  & \lambda\sqrt{H_{11}H_{22}} \cr \lambda\sqrt{H_{11}H_{22}} & \delta_2 H_{22}}.
\eequ{HiWi_matrices}
Note that $H_{11}$ and $H_{22}$ are real positive parameters defined similarly to those introduced by \citet{Suo1} for orthotropic bimaterials:
\beq
H_{11}=[bs_{11}^{'}]^{(1)}+[bs_{11}^{'}]^{(2)},\quad H_{22}=[es_{11}^{'}]^{(1)}+[es_{11}^{'}]^{(2)}.
\eequ{H11H22}
Regarding matrix $\BH$, two non-dimensional Dundurs-like parameters are defined \citep{Ting2,Suo1,MorRad1}:
\beq
\Ga=\fr{[ds_{11}^{'}]^{(1)}+[ds_{11}^{'}]^{(2)}}{\sqrt{H_{11}H_{22}}},\quad \beta=\fr{[s_{11}^{'}c-s_{12}^{'}]^{(1)}-[s_{11}^{'}c-s_{12}^{'}]^{(2)}}{\sqrt{H_{11}H_{22}}},
\eequ{Dundurs_sym}
while the matrix $\BW$ depends by four non-dimensional Dundurs-like parameters \citep{Ting2,Suo1,MorRad1}:
\beq
\delta_1=\fr{[bs_{11}^{'}]^{(1)}-[bs_{11}^{'}]^{(2)}}{H_{11}},\quad \delta_2=\fr{[es_{11}^{'}]^{(1)}-[es_{11}^{'}]^{(2)}}{H_{22}},
\eequ{Dundurs_skew1}
\beq
\lambda=\fr{[ds_{11}^{'}]^{(1)}-[ds_{11}^{'}]^{(2)}}{\sqrt{H_{11}H_{22}}},\quad \gamma=-\fr{[s_{11}^{'}c-s_{12}^{'}]^{(1)}+[s_{11}^{'}c-s_{12}^{'}]^{(2)}}{\sqrt{H_{11}H_{22}}}.
\eequ{Dundurs_skew2}
The Fourier transforms of the symmetric and skew-symmetric weight functions \eq{UsymH} and \eq{UskewW} for a plane monoclinic bimaterial assume the form:
\beq
\jump{0.1}{\hat{\BU}}^+(\xi) = -\fr{1}{|\xi|}\left(\BH^{'}-i\mbox{sign}(\xi)\beta\sqrt{H_{11}H_{22}}\BE\right)\hat{\BGS}^-(\xi),
\eequ{weightplane_sym}
\beq
\langle \hat{\BU} \rangle(\xi) = -\fr{1}{2|\xi|}\left(\BW^{'}+i\mbox{sign}(\xi)\gamma\sqrt{H_{11}H_{22}}\BE\right)\hat{\BGS}^-(\xi).
\eequ{weightplane_skew}
Since in plane strain elastic bimaterials Mode I and Mode II are coupled, two linearly independent singular solutions and tractions $\BU^i=(U_1^i,U_2^i)^T, \quad  \BGS^i=(\GS_{21}^i,\GS_{22}^i)^T, i=1,2,$ are needed in 
order to define a complete basis of the singular solutions space \citep{PiccMish1}. As a consequence, in this case symmetric and skew-symmetric weight functions [\BU] and $\langle \BU \rangle$, and the associate traction 
$\BGS$ are represented by $2\times2$ tensors which may be constructed by ordering the components of each singular solution in columns:
\beq
\BU=\pmatrix{U^1_1 & U^2_1 \cr U^1_2 & U^2_2}, \quad \BGS=\pmatrix{\GS^1_{21} & \GS^2_{21} \cr \GS^1_{22} & \GS^2_{22}}.
\eequ{USig_matrices}
Correspondingly, the rotation matrix $\tilde{\BR}$ reduces to:
\beq
\tilde{\BR}=\pmatrix{-1 & 0 \cr 0 & 1}.
\eequ{R_matrix}
Applying the Fourier transform to the \eq{Betti_conv}, we obtain:
\beq
\jump{0.1}{\hat{\BU}}^{T}\tilde{\BR}\hat{\bfm\tau}^{+}-\hat{\BGS}^{T}\tilde{\BR}\jump{0.1}{\hat{\Bu}}^{-}=-\jump{0.1}{\hat{\BU}}^{T}\tilde{\BR}\langle\hat{\Bp}\rangle-\langle\hat{\BU}\rangle^{T}\tilde{\BR}\jump{0.1}
{\hat{\Bp}},\quad \xi\in \textrm{R}.
\eequ{Betti_planefourier}
Multiplying both sides by $\tilde{\BR}^{-1}\jump{0.1}{\hat{\BU}}^{-T}$, the following identity is derived:
\beq
\hat{\bfm\tau}^{+}-\BB\jump{0.1}{\hat{\Bu}}^{-}=-\left\langle\hat{\Bp}\right\rangle-\BA\jump{0.1}{\hat{\Bp}},
\eequ{int_plane_fourier}
where $\BA$ and $\BB$ are given by:
\beq
\BA=\tilde{\BR}^{-1}\jump{0.1}{\hat{\BU}}^{-T}\langle \hat{\BU}^{T} \rangle\tilde{\BR}, \quad \BB=\tilde{\BR}^{-1}\jump{0.1}{\hat{\BU}}^{-T}\hat{\BGS}^{T}\tilde{\BR}.
\eequ{AB_plane}
Explicit expressions for these matrices can be computed using symmetric and skew-symmetric weight functions \eq{weightplane_sym} and \eq{weightplane_skew}:
\beq
\BA=\fr{1}{2\sqrt{H_{11}H_{22}}(\alpha^{2}+\beta^{2}-1)}(\BA^{'}+i\mbox{sign}(\xi)\BA^{''}),
\eequ{A_matrice}
\beq
\BB=\fr{|\xi|}{\sqrt{H_{11}H_{22}}(\alpha^{2}+\beta^{2}-1)}(\BB^{'}+i\beta\mbox{sign}(\xi)\BE),
\eequ{B_matrice}
where $\BA^{'}, \BA^{''}$ and $\BB^{'}$ are:
\beq
\BA^{'} = \pmatrix{\sqrt{H_{11}H_{22}}(\Ga\lambda-\Gb\gamma-\delta_1)  &  H_{22}(\lambda-\Ga\delta_2)  \cr  H_{11}(\lambda-\Ga\delta_1) & \sqrt{H_{11}H_{22}}(\Ga\lambda-\Gb\gamma-\delta_2)}, 
\eequ{Aimatrix}
\beq
\BA^{''} = \pmatrix{-\sqrt{H_{11}H_{22}}(\Ga\gamma+\Gb\lambda) & H_{22}(\gamma+\Gb\delta_2) \cr -H_{11}(\gamma+\Gb\delta_1) & \sqrt{H_{11}H_{22}}(\Ga\gamma+\Gb\lambda)},
\eequ{Aiimatrix}
\beq
\BB^{'} = \pmatrix{\sqrt{\fr{H_{22}}{H_{11}}} & \Ga \cr \Ga & \sqrt{\fr{H_{11}}{H_{22}}}}.
\eequ{Bimatrix}
Applying the inverse Fourier transform to the identity \eq{int_plane_fourier}, for the two cases $x_1<0$ and $x_1>0$, we get:
\beq
\left\langle \Bp \right\rangle(x_1) + \CF^{-1}_{x_{1}<0}\Big[\BA[\hat{\Bp}]\Big] = \CF^{-1}_{x_{1}<0}\Big[\BB[\hat{\Bu}]^{-}\Big], \quad x_1<0,
\eequ{int_plane1}
\beq
\bfm\tau^{(+)}(x_1) + \CF^{-1}_{x_{1}>0}\Big[\BA[\hat{\Bp}]\Big] = \CF^{-1}_{x_{1}>0}\Big[\BB[\hat{\Bu}]^{-}\Big], \quad x_1>0.
\eequ{int_plane2}
As for the case of antiplane deformations, illustrated in the previous Section, the term $\hat{\bfm\tau}^{+}$ in eq. (\ref{int_plane_fourier}) cancels from the \eq{int_plane1} because it is a ``$+$'' 
function, while $\langle\hat{\Bp}\rangle$ cancels from the \eq{int_plane2} because it is a ``$-$'' function.

Using the same inversion procedure of the previous Section the following integral identities for plane strain deformations in monoclinic bimaterials are derived:
\beq
\left\langle \Bp \right\rangle(x_1) + \BCA^{(s)}\jump{0.1}{\Bp} = \BCB^{(s)}\fr{\partial \jump{0.1}{\Bu}^{(-)}}{\partial x_1}, \quad x_1<0,
\eequ{int_plane1S}
\beq
\bfm\tau^{+}(x_1) + \BCA^{(c)}\jump{0.1}{\Bp} = \BCB^{(c)}\fr{\partial \jump{0.1}{\Bu}^{(-)}}{\partial x_1}, \quad x_1>0,
\eequ{int_plane2S}
where matrix operators $\BCA^{(s)}, \BCB^{(s)}: F(\textrm{R}_-)\rightarrow F(\textrm{R}_-)$, and $\BCA^{(c)}, \BCB^{(c)}: F(\textrm{R}_-)\rightarrow F(\textrm{R}_+)$ are defined as follows:
\beq
\BCA^{(s)}=\fr{1}{2\sqrt{H_{11}H_{22}}(\alpha^{2}+\beta^{2}-1)}(\BA^{'}+\BA^{''}\CS^{(s)}),
\eequ{As_operator}
\beq
\BCB^{(s)}=\fr{1}{\sqrt{H_{11}H_{22}}(\alpha^{2}+\beta^{2}-1)}(\BB^{'}\CS^{(s)}-\beta\BE),
\eequ{Bs_operator}
\beq
\BCA^{(c)}=\fr{1}{2\sqrt{H_{11}H_{22}}(\alpha^{2}+\beta^{2}-1)}\BA^{''}\CS^{(c)},
\eequ{Ac_operator}
\beq
\BCB^{(c)}=\fr{1}{\sqrt{H_{11}H_{22}}(\alpha^{2}+\beta^{2}-1)}\BB^{'}\CS^{(c)}.
\eequ{Bc_operator}
Equations \eq{int_plane1S} \eq{int_plane2S}, together with the definition of operators \eq{As_operator}, \eq{Bs_operator}, \eq{Ac_operator} and \eq{Bc_operator}, form the system of integral identities for Mode I and II 
deformations in monoclinic bimaterials. The equation \eq{int_plane1S} is a system of two coupled singular integral equations, which decouples in the case where the Dundurs parameters $\alpha$ and $\beta$ vanish. 
Observing expression 
\eq{Dundurs_sym}, we can note that $\beta$ vanishes in the case of a homogeneous monoclinic material, while $\alpha$ is zero only for some particular subclasses of materials, such as for orthotropic 
materials, where the quantity $d$, defined by \eq{mu_sumprod} and representing the imaginary part of the product of the eigenvalues, vanishes \citep{Suo2,GupArg1}. As a consequence, for a homogeneous 
orthotropic material, the system \eq{int_plane1S} is reduced to the following decoupled equations:
\begin{eqnarray}
-\fr{1}{H_{11}}\CS^{(s)}\fr{\partial \jump{0.1}{u_1}^{(-)}}{\partial x_1} & = & \left\langle p_1 \right\rangle(x_1) - \fr{\gamma}{2}\sqrt{\fr{H_{22}}{H_{11}}}\CS^{(s)}\jump{0.1}{p_2},\quad x_1<0,\\
-\fr{1}{H_{22}}\CS^{(s)}\fr{\partial \jump{0.1}{u_2}^{(-)}}{\partial x_1} & = & \left\langle p_2 \right\rangle(x_1) - \fr{\gamma}{2}\sqrt{\fr{H_{11}}{H_{22}}}\CS^{(s)}\jump{0.1}{p_1},\quad x_1<0.
\end{eqnarray}
The solution of these two equations requires the inversion of the singular operator $\CS^{(s)}$, which has been performed and discussed in details by \citet{PiccMish3}. The inversion of the matrix operator $\BCB^{(s)}$ 
in the general case requires the analysis of the systems of singular integral differential equations \citep{Vekua1}.

\section{Illustrative examples: line forces applied at the crack faces}
\label{examples}

\begin{figure}[!htcb]
\centering
\includegraphics[width=140mm]{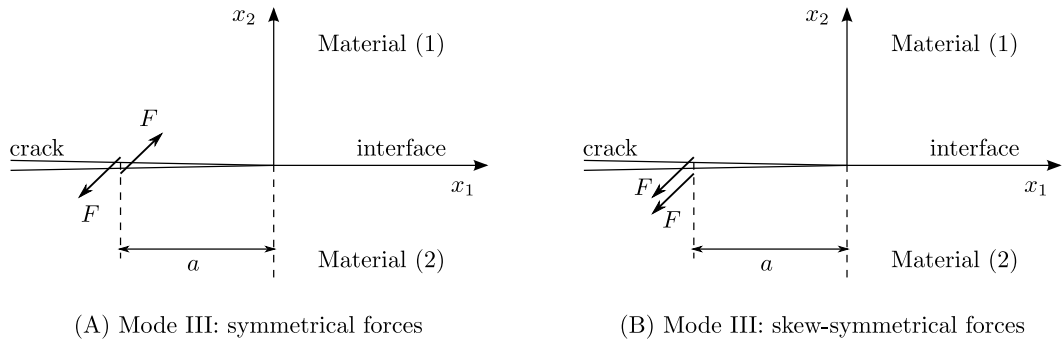}
\caption[]{\footnotesize Geometry of the Mode III crack.}
\label{fig02a}
\end{figure}

\begin{figure}[!htcb]
\centering
\includegraphics[width=140mm]{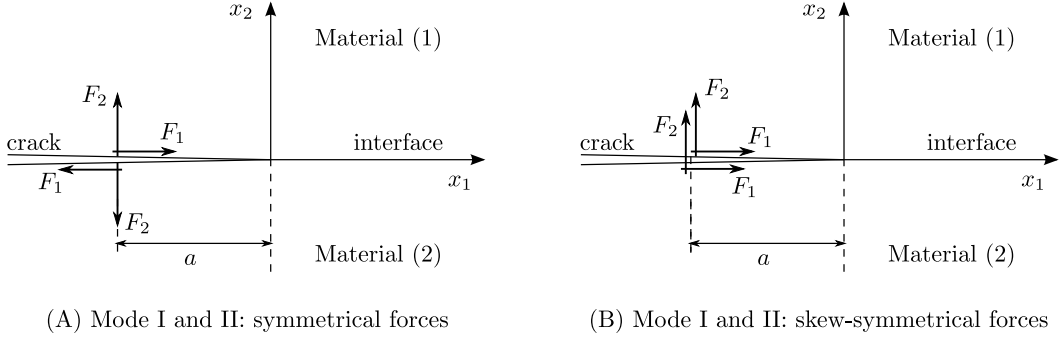}
\caption[]{\footnotesize Geometry of the Mode I and II crack.}
\label{fig02b}
\end{figure}

In this section we report an illustrative example of application of the integral identities in analysis of interfacial cracks in anisotropic bimaterials. Antiplane (Fig. \ref{fig02a}) and plane strain 
(Fig. \ref{fig02b}) interfacial cracks in monoclinic bimaterials loaded by line forces acting on the crack faces are studied by means of the proposed integral formulation. Explicit expressions for crack opening and 
tractions ahead of the tip corresponding to both symmetrical and skew-symmetrical loading configurations are derived. The proposed illustrative cases show that the integral identities derived in previous Section 
represent a very useful tool for studying interfacial crack problems in anisotropic materials, and their relevance is even greater in analysis of scenarios where the elastic problem is coupled with other phenomena, 
as for example hydraulic fracturing where both anisotropic elastic behaviour of geomaterials and fluid motion must be taken into account.

\vspace{3mm}

\subsection{Mode III: symmetrical line forces}

We consider an antiplane shear crack where the loading is given by two symmetrical line forces applied on the faces, at a distance $a$ from the crack tip, and directed along the $x_3$-axis:
\beq
\left\langle p_3 \right\rangle(x_1)=-F\delta(x_1+a),\quad \jump{0.1}{p_3}(x_1)=0,
\eequ{anti_symload}
where $\delta$ is the Dirac delta function, and F has the dimensions of a force divided by a length.

For antiplane deformations, the singular integral equations relating the applied loading and the resulting crack opening is given by \eq{int_anti1S}. Inverting the operator $\CS^{(s)}$ by means of procedure proposed by 
\citet{PiccMish3}, we obtain:
\beq
\fr{\partial \jump{0.1}{u_3}^{(-)}}{\partial x_1}=-\fr{H_{33}}{\pi}F\int_{-\infty}^0 \sqrt{\fr{\eta}{x_1}}\fr{\delta(\eta+a)}{x_1-\eta}d\eta=-\fr{H_{33}}{\pi}F\sqrt{-\fr{a}{x_1}}\fr{1}{x_1+a}.
\eequ{anti_jumpdevsym}
Assuming the condition that the crack opening vanishes at zero and at infinity, and integrating the \eq{anti_jumpdevsym}, the following expressions are derived:
\begin{eqnarray}
\jump{0.1}{u_3}(x_1) & = & \fr{2F}{\pi}H_{33}\, \mbox{arctanh}\sqrt{-\fr{x_1}{a}}, \quad -a < x_1 < 0,\nonumber\\
\jump{0.1}{u_3}(x_1) & = & \fr{2F}{\pi}H_{33}\, \mbox{arctanh}\sqrt{-\fr{a}{x_1}}, \quad x_1 < -a.
\label{anti_jumpsym}
\end{eqnarray}
Substituting \eq{anti_jumpdevsym} into equation \eq{intid_anti2}, the explicit expression for the traction ahead of the crack tip can be evaluated:
\beq
\tau_{3}^{(+)}(x_1)=-\fr{1}{\pi H_{33}}\int_{-\infty}^0 \fr{1}{x_1-\eta}\fr{\partial \jump{0.1}{u_3}^{-}}{\partial \eta} d\eta=\fr{F}{\pi}\sqrt{\fr{a}{x_1}}\fr{1}{x_1+a}.
\eequ{anti_tracsym}

It is important to note that the derived expressions for the traction ahead of the crack tip, and the associated crack opening, are identical to the results derived by \citet{PiccMish3} for the isotropic case, 
except for the parameter $H_{33}$, that is related to the anisotropy of the media. 

The traction expression \eq{anti_tracsym} can then be used for calculating the stress intensity factor. Applying to this specific case the general definition introduced for interfacial cracks in anisotropic 
materials by \cite{Wu1} and \cite{Hwu3} we obtain:
\beq
K_{III}=\lim_{x_1\rightarrow 0}\sqrt{2\pi x_1}\tau_{3}^{(+)}(x_1)=\sqrt{\fr{2}{\pi a}}F.
\eequ{Kfac_anti_sym}
The stress intensity factor \eq{Kfac_anti_sym}, calculated using \eq{anti_tracsym}, is identical to that obtained by \cite{Hwu3} for the same symmetric loading configuration applying a different procedure. This is an 
important proof for the validity of the traction expression \eq{anti_tracsym} and consequently of the associated crack opening \eq{anti_jumpsym}, derived by means of integral identities.

\subsection{Mode III: skew-symmetrical line forces}

We now consider an antiplane shear crack where the loading is given by two skew-symmetrical line forces applied on the faces, at a distance $a$ behind the crack tip, and directed along the $x_3$-axis:
\beq
\left\langle p_3 \right\rangle(x_1)=0,\quad \jump{0.1}{p_3}(x_1)=-2F\delta(x_1+a).
\eequ{anti_skewload}

Applying the inverse operator $(\CS^{(s)})^{-1}$ to equation \eq{int_anti1S}, we obtain: 
\beq
\fr{\partial \jump{0.1}{u_3}^{(-)}}{\partial x_1}=-\nu\fr{H_{33}}{\pi}F\int_{-\infty}^0 \sqrt{\fr{\eta}{x_1}}\fr{\delta(\eta+a)}{x_1-\eta}d\eta=-\nu\fr{H_{33}}{\pi}F\sqrt{-\fr{a}{x_1}}\fr{1}{x_1+a}.
\eequ{anti_jumpdevskew}
Integrating this expression the crack opening is then derived:
\begin{eqnarray}
\jump{0.1}{u_3}(x_1) & = & \nu\fr{2F}{\pi}H_{33}\, \mbox{arctanh}\sqrt{-\fr{x_1}{a}}, \quad -a< x_1 <0,\nonumber\\
\jump{0.1}{u_3}(x_1) & = & \nu\fr{2F}{\pi}H_{33}\, \mbox{arctanh}\sqrt{-\fr{a}{x_1}}, \quad x_1 < -a.
\label{anti_jumpskew}
\end{eqnarray}
Substituting \eq{anti_jumpdevskew} into equation \eq{intid_anti2}, the traction ahead of the crack tip becomes:
\beq
\tau_{3}^{(+)}(x_1)=\nu\fr{F}{\pi}\sqrt{\fr{a}{x_1}}\fr{1}{x_1+a}.
\eequ{anti_tracskew}

The derived expressions (\ref{anti_jumpskew}), \eq{anti_tracskew} are consistent with the results obtained in \citet{PiccMish3} for isotropic media, the only difference consists in the parameter $\nu$ which characterizes 
antiplane deformations in anisotropic bimaterials.

Also in this case, since traction expression \eq{anti_tracskew} have been derived, the stress intensity factor can be evaluated using the general definition:
\beq
K_{III}=\lim_{x_1\rightarrow 0}\sqrt{2\pi x_1}\tau_{3}^{(+)}(x_1)=\nu\sqrt{\fr{2}{\pi a}}F.
\eequ{Kfac_anti_skew}

The explicit expressions for the crack opening (\ref{anti_jumpsym}) and (\ref{anti_jumpskew}), and for the tractions ahead of the tip \eq{anti_tracsym} and \eq{anti_tracskew} have been derived only by inversion of the 
operator $\CS^{(s)}$ and by a simple integration procedure. This example shows that the integral identities obtained in Section \ref{Monoclinic} are fairly easy to use for solving antiplane interface crack problems in 
anisotropic materials. Then the integral identities represent a powerful tool for solving many problems in linear elastic fracture mechanics, especially for studying systems where the elastic fields are coupled with 
other physical phenomena.

\subsection{Mode I and II: symmetrical line forces}
\label{plane_strain}

The plane strain crack problem is now addressed. We assume that the loading is given by two symmetrical line forces applied on the faces at a distance $a$ from the crack tip and directed respectively along $x_1$ and 
$x_2$ axis:
\begin{eqnarray}
\left\langle p_1 \right\rangle(x_1) & = & -F_1\delta(x_1+a),\quad \jump{0.1}{p_1}(x_1)=0,\\
\left\langle p_2 \right\rangle(x_1) & = & -F_2\delta(x_1+a),\quad \jump{0.1}{p_2}(x_1)=0,
\end{eqnarray}
where $F_1$ and $F_2$ have the dimensions of a force divided by a length.

In two-dimensional interfacial elastic crack problems (plane strain or plane stress), Mode I and II are in general coupled and stress oscillations near the crack tip must be taken into account \citep{Suo1, Hwu3}. The 
corresponding integral identities \eq{int_plane1S} are a system of equations coupled by means of the generalized Dundurs parameter $\beta$ (see equation \eq{Dundurs_sym}), connected to the oscillation index of the 
bimaterial $\varepsilon$ \citep{Suo1}:
\beq
\varepsilon=\fr{1}{2\pi}\ln\left(\fr{1-\beta}{1+\beta}\right).
\eequ{osc}
For simplicity, we assume that $\beta$ is zero, this implies that also the oscillation index \eq{osc} vanishes and that in this particular case we have no oscillations at the crack tip. The equations \eq{int_plane1S} 
for $x_1<0$ become:
\begin{eqnarray}
\left\langle p_1 \right\rangle(x_1) & = & \fr{1}{\sqrt{H_{11}H_{22}}(\alpha^2-1)}
\left[\sqrt{\fr{H_{22}}{H_{11}}}\CS^{(s)}\fr{\partial \jump{0.1}{u_1}^{(-)}}{\partial x_1}+\alpha\CS^{(s)}\fr{\partial \jump{0.1}{u_2}^{(-)}}{\partial x_1}\right],\label{eqw1} \\
\left\langle p_2 \right\rangle(x_1) & = & \fr{1}{\sqrt{H_{11}H_{22}}(\alpha^2-1)}
\left[\alpha\CS^{(s)}\fr{\partial \jump{0.1}{u_1}^{(-)}}{\partial x_1}+\sqrt{\fr{H_{11}}{H_{22}}}\CS^{(s)}\fr{\partial \jump{0.1}{u_2}^{(-)}}{\partial x_1}\right].\label{eqw2}
\end{eqnarray}
Applying the inverse operator $(\CS^{(s)})^{-1}$ to both these equations, by means of some algebraic manipulations the following result is obtained:
\begin{eqnarray}
\fr{\partial \jump{0.1}{u_1}^{(-)}}{\partial x_1} & = & -\fr{H_{11}}{\pi}\left(F_1-\alpha\sqrt{\fr{H_{22}}{H_{11}}}F_2\right)\sqrt{-\fr{a}{x_1}}\fr{1}{x_1+a}, \\
\fr{\partial \jump{0.1}{u_2}^{(-)}}{\partial x_1} & = & -\fr{H_{22}}{\pi}\left(F_2-\alpha\sqrt{\fr{H_{11}}{H_{22}}}F_1\right)\sqrt{-\fr{a}{x_1}}\fr{1}{x_1+a}. 
\end{eqnarray}
Then, after integration, for $-a<x_1<0$ we derive:
\begin{eqnarray}
\jump{0.1}{u_1}^{(-)}(x_1)& = & \fr{2H_{11}}{\pi}\left(F_1-\alpha\sqrt{\fr{H_{22}}{H_{11}}}F_2\right)\mbox{arctanh}\sqrt{-\fr{x_1}{a}},\nonumber\\
\jump{0.1}{u_2}^{(-)}(x_1)& = & \fr{2H_{22}}{\pi}\left(F_2-\alpha\sqrt{\fr{H_{11}}{H_{22}}}F_1\right)\mbox{arctanh}\sqrt{-\fr{x_1}{a}}.\label{jump_plane_sym1}
\end{eqnarray}
And for $x_1<-a$:
\begin{eqnarray}
\jump{0.1}{u_1}^{(-)}(x_1)& = & \fr{2H_{11}}{\pi}\left(F_1-\alpha\sqrt{\fr{H_{22}}{H_{11}}}F_2\right)\mbox{arctanh}\sqrt{-\fr{a}{x_1}},\nonumber\\
\jump{0.1}{u_2}^{(-)}(x_1)& = & \fr{2H_{22}}{\pi}\left(F_2-\alpha\sqrt{\fr{H_{11}}{H_{22}}}F_1\right)\mbox{arctanh}\sqrt{-\fr{a}{x_1}}.\label{jump_plane_sym2}
\end{eqnarray}
The tractions components ahead of the crack tip can be evaluated by \eq{int_plane2S}:
\begin{eqnarray}
\tau_1^{(+)}(x_1) & = & \fr{F_1}{\pi}\sqrt{\fr{a}{x_1}}\fr{1}{x_1+a},\label{tracsym_plane1}\\
\tau_2^{(+)}(x_1) & = & \fr{F_2}{\pi}\sqrt{\fr{a}{x_1}}\fr{1}{x_1+a}.\label{tracsym_plane2}
\end{eqnarray}
We can observe that, also for the two-dimensional vector problem, if the loading is given by symmetric line forces applied at the faces, the corresponding expressions for the traction components ahead of 
the crack tip and for stress intensity factors are analogous to that derived by \citet{PiccMish3} for isotropic bimaterials, the only difference consists in the constants. The crack opening components 
(\ref{jump_plane_sym1}) and (\ref{jump_plane_sym2}) are coupled to $F_1$ and $F_2$ by means of the Dundurs parameter $\alpha$. This means that, even in the case where the symmetric load possesses only one non-zero 
component, directed along $x_1$-axis or $x_2$-axis, the induced crack opening has both $x_1$ and $x_2$ components. This aspect is an important difference respect to the case of isotropic materials, and it is connected 
with anisotropic properties of monoclinic media. 

In the particular case where $\alpha=0$, corresponding to orthotropic bimaterials, \citep{Suo2,GupArg1}, crack opening expressions are similar to the isotropic case, except for the different constants. For $-a<x_1<0$:
\begin{eqnarray}
\jump{0.1}{u_1}^{(-)}(x_1)& = & \fr{2H_{11}}{\pi}F_1\mbox{arctanh}\sqrt{-\fr{x_1}{a}},\nonumber\\
\jump{0.1}{u_2}^{(-)}(x_1)& = & \fr{2H_{22}}{\pi}F_2\mbox{arctanh}\sqrt{-\fr{x_1}{a}}.\label{jump_plane_symortho1}
\end{eqnarray}
And for $x_1<-a$:
\begin{eqnarray}
\jump{0.1}{u_1}^{(-)}(x_1)& = & \fr{2H_{11}}{\pi}F_1\mbox{arctanh}\sqrt{-\fr{a}{x_1}},\nonumber\\
\jump{0.1}{u_2}^{(-)}(x_1)& = & \fr{2H_{22}}{\pi}F_2\mbox{arctanh}\sqrt{-\fr{a}{x_1}}.\label{jump_plane_symortho2}
\end{eqnarray}

The tractions expressions (\ref{tracsym_plane1}) and (\ref{tracsym_plane2}) can then be used for evaluating the stress intensity factors. Applying the general definition explained in details in \cite{Wu1} and 
\cite{Hwu3} to this particular two-dimensional case without oscillation, we obtain the vectorial formula:
\beq
\BK=\lim_{x_1\rightarrow 0}\sqrt{2\pi x_1}\bfm\tau^{(+)}(x_1),
\eequ{sif_plane}
where $\BK=(K_{II},K_I)^T$ and $\bfm\tau^{(+)}=(\tau^{(+)}_{1},\tau^{(+)}_{2})^T$. Inserting (\ref{tracsym_plane1}) and (\ref{tracsym_plane2}) and evaluating the limit, we get:
\beq
K_{I}=\sqrt{\fr{2}{\pi a}}F_2, \quad K_{II}=\sqrt{\fr{2}{\pi a}}F_1.
\eequ{sif_planesym}

Also in this case, as is it has been detected for the antiplane problem, the derived stress intensity factors are equal to that obtained by \cite{Hwu3} in the limit of vanishing oscillatory index. This demonstrates that 
the expressions (\ref{tracsym_plane1}) and (\ref{tracsym_plane2}), obtained by means of the integral identities, are correct.

In several studies, a unique complex stress intensity factor accounting both Mode I and Mode II contributions is associated to two-dimensional interfacial crack problems in anisotropic materials \citep{Suo1, SuoKuo1, 
MorRad1}, similarly to what happens due to the oscillations for cracks between different isotropic media \citep{MantPar1, PiccMish2, GracMant1}. It has been shown by \cite{Hwu3} that complex stress intensity factor, 
introduced for anisotropic bimaterials by \cite{Suo1}, is linked to the classical stress intensity factors by a vectorial transformation. In our example, we have assumed that the generalized Dundurs parameter $\beta$ and 
the correlated oscillatory index are zero. In this particular case, where there are no oscillations at the interface and Mode I and II can be decoupled, \cite{Suo1} and \cite{Hwu3} analyses lead to the same expressions 
\eq{sif_plane} for real stress intensity factors.

\subsection{Mode I and II: skew-symmetrical line forces}

We assume that the loading is given by two skew-symmetrical line forces applied on the faces at a distance $a$ from the crack tip and directed respectively along $x_1$ and $x_2$ axis:
\begin{eqnarray}
\left\langle p_1 \right\rangle(x_1) & = & 0,\quad \jump{0.1}{p_1}(x_1)=-2F_1\delta(x_1+a),\\
\left\langle p_2 \right\rangle(x_1) & = & 0,\quad \jump{0.1}{p_2}(x_1)=-2F_2\delta(x_1+a).
\end{eqnarray}
Also in this case we consider the case where the Dundurs parameter $\beta$ associated to crack tip oscillations vanishes. Furthermore, also $\lambda$ is assumed to be zero. Applying the inverse operator 
$(\CS^{(s)})^{-1}$ to the \eq{int_plane1S} for $x_1<0$ we obtain:
\begin{eqnarray}
\fr{\partial \jump{0.1}{u_1}^{(-)}}{\partial x_1} & = & -\fr{F_1}{\pi}\delta_1 H_{11}\sqrt{-\fr{a}{x_1}}\fr{1}{x_1+a}-\gamma F_2 \sqrt{H_{11}H_{22}}\delta(x_1+a),\\
\fr{\partial \jump{0.1}{u_2}^{(-)}}{\partial x_1} & = & -\fr{F_2}{\pi}\delta_2 H_{22}\sqrt{-\fr{a}{x_1}}\fr{1}{x_1+a}+\gamma F_1 \sqrt{H_{11}H_{22}}\delta(x_1+a). 
\end{eqnarray}
Integrating these expressions, for $-a<x_1<0$ we derive:
\begin{eqnarray}
\jump{0.1}{u_1}^{(-)}(x_1)& = & \fr{2F_1}{\pi}\delta_1 H_{11}\mbox{arctanh}\sqrt{-\fr{x_1}{a}},\nonumber\\
\jump{0.1}{u_2}^{(-)}(x_1)& = & \fr{2F_2}{\pi}\delta_2 H_{22}\mbox{arctanh}\sqrt{-\fr{x_1}{a}},\label{jump_plane_skew1}
\end{eqnarray}
And for $x_1<-a$:
\begin{eqnarray}
\jump{0.1}{u_1}^{(-)}(x_1)& = & \fr{2F_1}{\pi}\delta_1 H_{11}\mbox{arctanh}\sqrt{-\fr{a}{x_1}}+\gamma F_2 \sqrt{H_{11}H_{22}},\nonumber\\
\jump{0.1}{u_2}^{(-)}(x_1)& = & \fr{2F_2}{\pi}\delta_2 H_{22}\mbox{arctanh}\sqrt{-\fr{a}{x_1}}-\gamma F_1 \sqrt{H_{11}H_{22}},\label{jump_plane_skew2}
\end{eqnarray}
Except for the constants, these expressions for the crack opening are identical to those obtained by \citet{PiccMish3} for the isotropic materials and benchmarked by a comparison with the Flamant solution for a 
half-plane loaded by two concentrated forces at its surface \citep{Barb1}. The tractions components ahead of the crack tip become:
\begin{eqnarray}
\tau_1^{(+)}(x_1) & = & \fr{1}{\pi (1-\alpha^2)}\left(\delta_1F_1+\alpha\delta_2 F_2\sqrt{\fr{H_{22}}{H_{11}}}\right)\sqrt{\fr{a}{x_1}}\fr{1}{x_1+a},\label{tract_plane_skew1}\\
\tau_2^{(+)}(x_1) & = & \fr{1}{\pi (1-\alpha^2)}\left(\delta_2F_2+\alpha\delta_1 F_1\sqrt{\fr{H_{11}}{H_{22}}}\right)\sqrt{\fr{a}{x_1}}\fr{1}{x_1+a}. \label{tract_plane_skew2}
\end{eqnarray}
From these expressions we can observe that, if the loading is given by skew-symmetric line forces applied at the faces, the corresponding traction components (\ref{tract_plane_skew1}) and (\ref{tract_plane_skew2}) are 
coupled to $F_1$ and $F_2$ by means of the Dundurs parameter $\alpha$. As a consequence, similarly to what we have detected for the jump behind the tip in presence of symmetric loading, even in the case where the skew-
symmetric load possesses only one non-zero component, directed along $x_1$-axis or $x_2$-axis, the associated tractions have both $x_1$ and $x_2$ components. This coupling effect is not found in isotropic materials, and 
it is connected with anisotropic properties of monoclinic media. 

\begin{figure}[htbp]
\centering
\includegraphics[width=15.7cm]{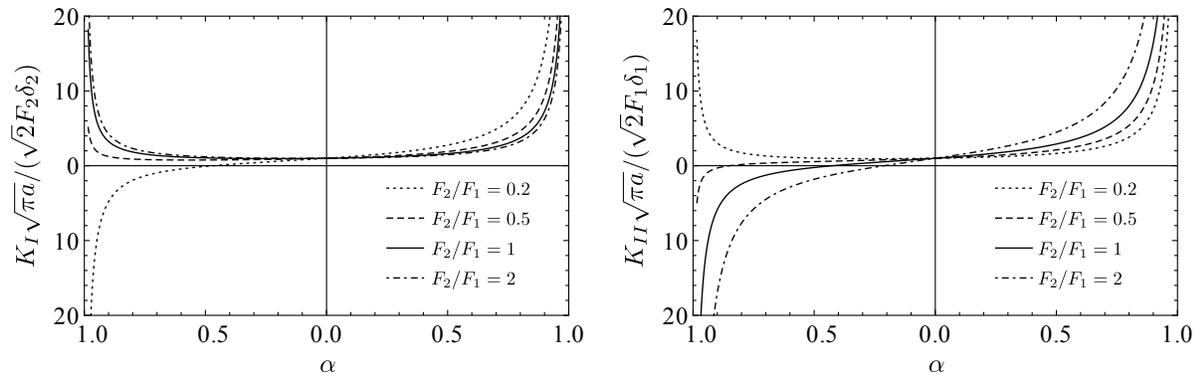}
\caption[Kfac_F]{\footnotesize {Normalized stress intensity factors corresponding to skew-symmetric forces (see Fig. \ref{fig02b}B) as a function of the bimaterial parameter $\alpha$ for 
$-0.9839\leq\alpha\leq0.9839$, computed for $H_{11}=2.01, H_{22}=6.98, \delta_1=0.72, \delta_2=0.92$ and different values of the ratio $F_2/F_1$: $F_2/F_1=0.2, F_2/F_1=0.5, F_2/F_1=1, F_2/F_1=2$.}}
\label{Kafc_F}
\end{figure}

The stress intensity factors can then be evaluated substituting the tractions (\ref{tract_plane_skew1}) and (\ref{tract_plane_skew2}) into the general expression \eq{sif_plane}:
\begin{eqnarray}
K_{I}  & = & \fr{1}{1-\alpha^2}\sqrt{\fr{2}{\pi a}}\left(\delta_2F_2+\alpha\delta_1 F_1\sqrt{\fr{H_{11}}{H_{22}}}\right),\label{sif_plane_skew1}\\
K_{II} & = & \fr{1}{1-\alpha^2}\sqrt{\fr{2}{\pi a}}\left(\delta_1F_1+\alpha\delta_2 F_2\sqrt{\fr{H_{22}}{H_{11}}}\right). \label{sif_plane_skew2}
\end{eqnarray}
For $\alpha = 0$, the materials assume orthotropic behaviour, and we recover tractions expressions similar to those obtained for isotropic media:
\begin{eqnarray}
\tau_1^{(+)}(x_1) & = & \fr{F_1}{\pi}\delta_1\sqrt{\fr{a}{x_1}}\fr{1}{x_1+a},\\
\tau_2^{(+)}(x_1) & = & \fr{F_2}{\pi}\delta_2\sqrt{\fr{a}{x_1}}\fr{1}{x_1+a}. 
\end{eqnarray}
The stress intensity factors become:
\beq
K_{I}=\delta_2\sqrt{\fr{2}{\pi a}}F_2, \quad K_{II}=\delta_1\sqrt{\fr{2}{\pi a}}F_1. 
\eequ{Kfac_plane_skew}
These expressions for $K_{I}$ and $K_{II}$, corresponding to orthotropic bimaterial, possess the same form obtained by \cite{PiccMish3} for isotropic media. \\
The skew-symmetric stress intensity factors (\ref{sif_plane_skew1}) and (\ref{sif_plane_skew2}) have been normalized, multiplying respectively by $\sqrt{\pi a}/(\sqrt{2} F_2 \delta_2)$ and  
$\sqrt{\pi a}/(\sqrt{2} F_1 \delta_1)$, and plotted in Fig. \ref{Kafc_F} as functions of the Dundurs-like parameter $\alpha$. In order to satisfy the positive definiteness of the bimaterial matrix 
$\BY$, and consequently to verify relations \eq{Dund_lim} \citep{Ting2}, the value of $\alpha$ must be included in an interval $\alpha_{\mbox{min}} \leq \alpha \leq \alpha_{\mbox{max}}$, where the bounds depend 
on the parameters $H_{11}, H_{22}, \delta_1$ and $\delta_2$. Assuming $H_{11} = 2.01, H_{22} = 6.98, \delta_1 = 0.72, \delta_2 = 0.92$, which corresponds approximately to a system where material 
$(1)$ is boron and material $(2)$ is aluminium \citep{Suo2}, we get $-0.9839\leq\alpha\leq0.9839$. The normalized stress intensity factors computed for values of $\alpha$ within this 
interval and for different values of the ratio $F_2/F_1$ between the applied forces are reported in Fig. \ref{Kafc_F}.
As it can be deduced by normalizing expressions \eq{Kfac_plane_skew}, for $\alpha=0$ the curves collapse and the materials assume orthotropic behaviour. On the other hand, as the value of $\alpha$ 
approaches the bounds $\alpha_{\mbox{max}} = 0.9839$ and $\alpha_{\mbox{min}} = -0.9839$, the absolute value of the skew-symmetric stress intensity factors increases and consequently the response of the system to 
skew-symmetric loading is amplified.

The examples illustrated in this Section and in Section \ref{plane_strain} show that the integral identities can be profitably used for solving vectorial interfacial crack problems in two-dimensional anisotropic elastic 
media. The crack opening and the tractions ahead of the tip can be derived without the use of Green's function, which implies challenging calculations. For this reason, also in the vectorial case, the proposed singular 
integral formulation may represent a very suitable technique in analysis of several fracture processes, especially where coupled fields are involved.

\section{Conclusions}

The problem of a two-dimensional semi-infinite interfacial crack in anisotropic bimaterials has been formulated in terms of singular integral equations by means of weight functions and integral transforms. The proposed 
method avoids the use of Green's function and the challenging numerical calculations related to such approach. Integral identities relating the applied loading and the corresponding crack opening have been obtained for 
both in-plane and antiplane strain problems in anisotropic bimaterials. Detailed explicit derivation of the identities have been performed for monoclinic materials, which are the most general class of 
anisotropic media allowing decoupling between both in-plane and antiplane stress and in-plane and antiplane strain \citep{Ting2,Ting3}. As a consequence, the obtained integral 
formulation can be directly used for studying cracks propagation in all anisotropic media where in-plane and antiplane problems are decoupled, such as cubic and orthotropic materials \citep{Suo1}. Thanks 
to the 
great versatility of the Stroh formalism, the developed approach can also be easily adapted for studying fracture phenomena in many other materials, such as piezoelectrics, poroelastics, and quasicrystals. 

An example of application of the integral identities to antiplane and plane strain crack problems where the loading is given by line forces acting on the faces has been performed. Explicit expressions for crack opening 
and tractions ahead of the tip corresponding to both symmetric and skew-symmetric loading configurations have been obtained.

The derived integral identities have various relevant applications especially to multifield problems, where the elasticity equations are coupled with other concurrent phenomena, for example, but not 
only, to hydraulic 
fracturing modelling. Furthermore, they also have their own value from the mathematical point of view, as, to the authors best knowledge, such identities written in a similar explicit form for interfacial cracks in 
anisotropic bimaterials seems to be unknown in the literature.

\section*{Ackowledgements:}  L. M. and E. R. gratefully thank financial support from the ''Cassa di Risparmio di Modena'' in the framework of the International Research Project $2009-2010$ "Modelling of crack propagation 
in complex materials". A.P. and G.M. gratefully acknowledge the support from the European Union Seventh Framework Programme under contract numbers PIEF-GA-2009-252857-INTERCRACKS and PIAP-GA-2011-286110-INTERCER2, 
respectively. 

\section*{A Anisotropic materials: Stroh formalism}
\label{Int_crack}

In this Appendix, complex variable representation for the stress field and displacements in anisotropic elastic materials subject to two-dimensional deformations based on Stroh formalism \citep{Stroh1} is summarized. 
Explicit expressions for Stroh matrices corresponding to monoclinic materials with symmetry plane at $x_3=0$ \citep{Ting2} are reported. The surface admittance tensor needed in the evaluation of bimaterial matrices 
\eq{UsymH} and \eq{UskewW} and then of weight functions (\ref{H}) and (\ref{W}) is derived.

\subsection*{A.1 Complex variable representation of stress and displacements}
\label{Stroh}
For two-dimensional problems in anisotropic elastic materials, displacements and stress fields can be represented in terms of complex variable functions matrices by means of two alternative formulations, proposed 
respectively by \citet{Stroh1} and \citet{Lekh1}. Introducing the stress vectors $\Bt_k=(\Gs_{1k},\Gs_{2k},\Gs_{3k} )^T,\:k=1,2$ and the displacements $\Bu=(u_1,u_2,u_3)^T$, the constitutive relations connecting the 
stresses and the strains are written using the Stroh formulation as follows:
\begin{eqnarray}
\Bt_1 & = & \BQ \Bu_{,1} + \BR \Bu_{,2},  \label{const1} \\
\Bt_2 & = & \BR^T {\Bu}_{,1} + \BT \Bu_{,2},\label{const2}
\end{eqnarray}
 The $3\times3$ matrices $\BQ, \BR$ and $\BT$ depend on the material constants, and are defined as follows \citep{Ting2}:
\beq
Q_{ik}=C_{i1k1},\quad R_{ik}=C_{i1k2}, \quad T_{ik}=C_{i2k2},
\eequ{matrices_stroh}
where $C_{ijkl}$ are components of the elastic stiffness tensor. Using this notation, the static equilibrium equations become \citep{Ting2, Ting3}:
\beq
\Bt_{1,1}+\Bt_{2,2}=\BQ \Bu_{,11}+(\BR+\BR^T) \Bu_{,12}+\BT \Bu_{,22}=0.
\eequ{equilibrium} 
The displacement $\Bu(x_{1},x_{2})$, which is a general solution of equation \eq{equilibrium}, has the following form \citep{Suo1,Ting1}:
\beq
\Bu=\BF \Bg(\Bz) + \ov{\BF \Bg(\Bz)},
\eequ{displac}
also the derivative of the displacements $\Bu_{,1}(x_{1},x_{2})$ and the traction $\Bt_2(x_{1},x_{2})$, can be written in the same form:
 \beq
 \Bt_2(x_{1},x_{2}) = \BL \Bh(\Bz) + \ov{\BL \Bh(\Bz)},
 \eequ{trac_stroh}
and
 \beq
 \Bu_{,1} (x_{1},x_{2}) = \BF \Bh(\Bz) + \ov{\BF \Bh(\Bz)},
 \eequ{displ_stroh}
 where $\Bh(\Bz)=d\Bg/d\Bz$, $\BF$ and $\BL$ are constant $3\times3$ matrices, defined as follows:
\beq
\BF=(\Bf_1,\Bf_2,\Bf_3), \quad \BL=(\Bl_1,\Bl_2,\Bl_3).
\eequ{AB_matrix}
Note that $\Bg(\Bz)$ and  $\Bh(\Bz)$ are analytic functions vectors with components $ g_j(x_{1}+ \mu_j x_{2})$, and $ h_j(x_{1}+ \mu_j x_{2})$ and $\mu_j$ are complex numbers with positive imaginary parts. According to 
\citet{Suo1}, if $g_{j}(z_{j})$ and $h_{j}(z_{j})$ are analytic functions of $z_{j}=x_{1}+ \mu_j x_{2}$ in the upper half-plane (or in the lower half-plane) for one $\mu_j$, where $\mu_j$ is a complex number with 
positive imaginary parts, they are analytic for any  $\mu_j$. On the basis of this property, here and in the text that follows, the analysis is reduced to a single complex variable. The eigenvectors $\Bf_j$ and the 
eigenvalues $\mu_j$ are simultaneously determined inserting expression \eq{displac} into equilibrium equations \eq{equilibrium}, so that they are reduced to the eigenvalue problem \citep{Ting1}:
\beq
[\BQ+(\BR+\BR^T)\mu_j+\BT\mu_j^{2}]\Bf_j = 0. 
\eequ{a_vec}
Moreover, $\Bl_j$ are related to $\Bf_j$ as follows \citep{Ting2,Ting1,Ting3}:
\beq
\Bl_j=[\BR^T + \mu_{j}\BT]\Bf_j.       
\eequ{eq4}  
The Hermitian surface admittance tensor, needed in general weight functions expressions \eq{Usym} and \eq{Uskew}, is defined as $\BY=i\BF\BL^{-1}$ \citep{GaoAbbu1}.

Stroh formalism have been extensively used in analysis of interfacial cracks in anisotropic bimaterials by \citet{Suo1} and \citet{GaoAbbu1}. Physical displacements and stress fields at the interface between the two 
materials have been derived, expressing the boundary conditions in terms of non-homogeneous Riemann-Hilbert problem and obtaining an algebraic eigenvalue involving the symmetric bimaterial matrix (\ref{H}), that is 
solved in closed form. Recently, this approach has been extended and used together with Fourier transform by \citet{MorRad1} for deriving singular solutions $\BU$ of the elasticity problem with zero traction on the faces 
where the crack is placed along the positive semi-axis $x_1>0$, and for evaluating general expressions for symmetric and skew-symmetric weight functions defined as traces of these functions, following the method 
illustrated in \citet{PiccMish1}. 

\subsection*{A.2 Monoclinic materials}
\label{stroh_monoclinic}

For monoclinic materials with the symmetry plane at $x_3=0$, employing the contracted notation of the stiffness tensor $C_{ijkl}$  \citep{Suo1,Ting2,Ting1}, the three matrices $\BQ,\BR$ and $\BT$ are given by:
$$
\BQ = \pmatrix{c_{11} & c_{16}  & 0 \cr c_{16} & c_{66} & 0 \cr 0 & 0 & c_{55}},\;
\BR = \pmatrix{c_{16} & c_{12} & 0 \cr c_{66} & c_{26} & 0 \cr 0 & 0 & c_{45}},\; \BT = \pmatrix{c_{66} & c_{26} & 0 \cr c_{26} & c_{22} & 0 \cr 0 & 0 & c_{44}} .
$$ 
This structure of the Stroh matrices is a consequence of the fact that for this class of anisotropic materials, the reduced $6\times6$ stiffness tensor $c_{ij}$ possesses the following important property 
\citep{Suo1,Ting3}:  
\beq
c_{14}=c_{15}=c_{24}=c_{25}=c_{46}=c_{56}=0.
\eequ{mon_const}
Substituting matrices $\BQ,\BR$ and $\BT$ into constitutive relations (\ref{const1}) and (\ref{const2}), we can easily observe that as a consequence of property \eq{mon_const}, for monoclinic materials subject to two-
dimensional deformations in-plane strain and antiplane strain are uncoupled and also in-plane stress antiplane stress are uncoupled. More discussions and details on decoupling of 
in-plane and antiplane deformations and stress are given by \citet{HorMil1} and \citet{Ting3}. The eigenvalues problem \eq{a_vec} then reduces to:
\beq
\pmatrix{c_{11}+2c_{16}\mu+c_{66}\mu^2 & c_{16}+(c_{12}+c_{66})\mu+c_{26}\mu^2 & 0\cr c_{16}+(c_{12}+c_{66})\mu+c_{26}\mu^2 & c_{66}+2c_{26}\mu+c_{22}\mu^2 & 0 \cr 0 & 0 & c_{55}+2c_{45}\mu+c_{44}\mu^2 }\Bf=0.
\eequ{eigenstroh}
Through the decoupling, the sixth order characteristic equation of this eigensystem consists in the product between a fourth order term corresponding to in-plane strain and a second order term associated 
to antiplane deformations \citep{Ting1}:
\beq
(c_{11}+2c_{16}\mu+c_{66}\mu^2)(c_{66}+2c_{26}\mu+c_{22}\mu^2)-(c_{16}+(c_{12}+c_{66})\mu+c_{26}\mu^2)^2=0,
\eequ{chareq_inplane} 
\beq
c_{55}+2c_{45}\mu+c_{44}\mu^2 =0.
\eequ{chareq_anti}
Since these equations possesses real coefficients, the roots are complex conjugates. Considering only eigenvalues with positive imaginary part \citep{Suo1}, $\mu_1$ and $\mu_2$ are assumed to be roots of equation 
\eq{chareq_inplane}, while $\mu_3$
is root of  \eq{chareq_anti}. 

The illustrated Stroh representation of the elasticity problem is equivalent to the matrix formulation proposed by \citet{Lekh1}, which provides alternative expressions for the eigenvector matrices $\BF$ and $\BL$ in 
function of the elements of the compliance matrix $\BS=\BC^{-1}$. More precisely, Lekhnitskii approach gives a specially normalized eigenvector matrix $\BF$, and expressing the elements of the stiffness matrix in 
function of the stiffness parameters it is easy to check that characteristic equation derived using Lekhnitskii formulation is identical to \eq{chareq_inplane} and \eq{chareq_anti} \citep{Suo1, Hwu1}. In order to obtain 
compact expressions for the surface admittance tensor $\BY$ and then for bimaterial matrices (\ref{H}) and (\ref{W}), particularly convenient for weight functions derivation, we assume this particular normalization for 
matrices $\BF$ and $\BL$, reported by \citet{Hwu1}, and we express the stiffness reduced tensor elements $c_{ij}$ in function of the elements of 
the reduced compliance matrix $s_{ij}^{'}$:
\beq
s^{'}_{ij}=s_{ij}-\frac{s_{i3}s_{3j}}{s_{33}}.
\eequ{red_compliance}
An alternative form for characteristic equations \eq{chareq_inplane} and \eq{chareq_anti} is derived:
\beq
s^{'}_{11}\mu^4-2s^{'}_{16}\mu^3+(2s^{'}_{12}+s^{'}_{66})\mu^2-2s^{'}_{26}\mu+s^{'}_{22}=0,
\eequ{chareq_plane_s}
\beq
s^{'}_{44}-2s^{'}_{45}\mu+s^{'}_{55}\mu^2 =0.
\eequ{chareq_anti_s}
The hermitian matrix $\BY=i\BF\BL^{-1}$ evaluated using the eigenvector normalization reported in \citet{Hwu1} assumes the form:
\beq
\BY = i\BF\BL^{-1} = 
i\pmatrix{s_{16}^{'}-s_{11}^{'}(\mu_1+\mu_2) & s_{12}^{'}-s_{11}^{'}\mu_1\mu_2 & 0 \cr 
\fr{s_{22}^{'}}{\mu_1\mu_2}-s_{12}^{'} & s_{22}^{'}\left( \fr{1}{\mu_1}+\fr{1}{\mu_2}\right)-s_{26}^{'} & 0 \cr 
0 & 0 & \fr{s^{'}_{44}}{\mu_3}-s_{45}^{'}},
\eequ{Ymatrix}
where $\mu_1$ and $\mu_2$ are roots of the equation \eq{chareq_plane_s}, and $\mu_3$ is solution of the equation \eq{chareq_anti_s} with positive imaginary part, corresponding respectively to plane strain and antiplane 
strain. Employing the relations between $\mu_1$ and $\mu_2$ and coefficients of \eq{chareq_plane_s} \citep{Suo1, Ting2} and between $\mu_3$ and coefficients of \eq{chareq_anti_s} \citep{Suo1}, through some manipulations 
the following form for $\BY$ is obtained:
\beq
\BY = i\BF\BL^{-1} = 
\pmatrix{s_{11}^{'}\mbox{Im}(\mu_1+\mu_2) & i(s_{12}^{'}-s_{11}^{'}\mu_1\mu_2) & 0 \cr 
i(s_{11}^{'}\ov{\mu}_1\ov{\mu}_2-s_{12}^{'}) & s_{11}^{'}\mbox{Im}(\mu_1\mu_2(\ov{\mu}_1+\ov{\mu}_2)) & 0 \cr 
0 & 0 & \sqrt{s_{44}^{'}s_{55}^{'}-s_{45}^{'2}}}.
\eequ{YmatrixLekh}
This general compact expression for the surface admittance tensor has been used for deriving explicit bimaterial matrices (\ref{H}) and (\ref{W}) and symmetric and skew-symmetric weight functions \eq{UsymH} and 
\eq{UskewW}. On the basis of property \eq{mon_const}, Mode III crack propagation (antiplane shear) will be treated separately from Mode I and Mode II (plane strain), as for the case of interfacial cracks in 
two-dimensional isotropic bimaterials.

\bibliography{WF_Bib} 

\bibliographystyle{apa-good}

\end{document}

%% file: macro.tex
\newcommand\eq[1] {(\ref{#1})}

\newcommand{\bfm}[1]{\mbox{\boldmath ${#1}$}}

\newcommand{\beqa}{\begin{eqnarray}}
\newcommand{\eeqa}[1]{\label{#1}\end{eqnarray}}
\newcommand{\bequ}{\begin{equation}}
\newcommand{\eequ}[1]{\label{#1}\end{equation}}

\newcommand{\ov}[1]{\overline{#1}}

\newcommand{\jump}[2]{[\mbox{\hspace{-#1em}}[#2]\mbox{\hspace{-#1em}}]}

\newcommand{\Ga}{\alpha}
\newcommand{\Gb}{\beta}

\newcommand{\Gs}{\sigma}

\newcommand{\GS}{\Sigma}

\newcommand{\BGs}{\bfm\sigma}

\newcommand{\BGS}{\bfm\Sigma}

\newcommand{\CF}{{\cal F}}

\newcommand{\CI}{{\cal I}}

\newcommand{\CP}{{\cal P}}

\newcommand{\CS}{{\cal S}}

\newcommand{\BCA}{{\bfm{\cal A}}}
\newcommand{\BCB}{{\bfm{\cal B}}}

\def\Ba{{\bf a}}
\def\Bb{{\bf b}}

\def\Bf{{\bf f}}
\def\Bg{{\bf g}}
\def\Bh{{\bf h}}

\def\Bl{{\bf l}}

\def\Bp{{\bf p}}

\def\Bt{{\bf t}}
\def\Bu{{\bf u}}

\def\Bz{{\bf z}}
\def\BA{{\bf A}}
\def\BB{{\bf B}}
\def\BC{{\bf C}}

\def\BE{{\bf E}}
\def\BF{{\bf F}}

\def\BH{{\bf H}}

\def\BK{{\bf K}}
\def\BL{{\bf L}}

\def\BP{{\bf P}}
\def\BQ{{\bf Q}}
\def\BR{{\bf R}}
\def\BS{{\bf S}}
\def\BT{{\bf T}}
\def\BU{{\bf U}}

\def\BW{{\bf W}}

\def\BY{{\bf Y}}

\newcommand{\beq}{\begin{equation}}
\newcommand{\eeq}{\end{equation}}
\newcommand{\overliner}{\begin{eqnarray}}
\newcommand{\earr}{\end{eqnarray}}
\newcommand{\beqn}{\begin{equation*}}
\newcommand{\eeqn}{\end{equation*}}
\newcommand{\overlinern}{\begin{eqnarray*}}
\newcommand{\earrn}{\end{eqnarray*}}

\newcommand{\fr}{\frac}

\def\f0{\ensuremath{\mathbb{O}}}